\documentclass{article}

     \PassOptionsToPackage{numbers, compress}{natbib}
    \usepackage[preprint]{neurips_2022_ml4ps}

\usepackage[utf8]{inputenc} % allow utf-8 input
\usepackage[T1]{fontenc}    % use 8-bit T1 fonts
\usepackage{hyperref}       % hyperlinks
\usepackage{url}            % simple URL typesetting
\usepackage{booktabs}       % professional-quality tables
\usepackage{amsfonts}       % blackboard math symbols
\usepackage{nicefrac}       % compact symbols for 1/2, etc.
\usepackage{microtype}      % microtypography
\usepackage{xcolor}         % colors
\usepackage{mathtools}
\usepackage{wrapfig}
\usepackage{amssymb}
\usepackage{amsthm}

\newcommand{\M}{\mathcal{M}}

\title{CaloMan: Fast generation of calorimeter showers \\ with  density estimation on learned manifolds}

\author{%
  Jesse C. Cresswell \\
  Layer 6 AI\\
  \texttt{jesse@layer6.ai} \\
  \And
  Brendan Leigh Ross \\
  Layer 6 AI\\
  \texttt{brendan@layer6.ai} \\
   \And
  Gabriel Loaiza-Ganem \\
  Layer 6 AI\\
  \texttt{gabriel@layer6.ai} \\
  \AND
  Humberto Reyes-Gonz\'alez \\
  University of Genoa \& INFN\\
\small\texttt{hreyes@ge.infn.it} \\
  \And
  Marco Letizia \\
  University Of Genoa \& INFN\\
  \texttt{marco.letizia@edu.unige.it} \\
  \And
  Anthony L. Caterini \\
  Layer 6 AI\\
  \texttt{anthony@layer6.ai} \\
}

\begin{document}

\maketitle

\begin{abstract}
    Precision measurements and new physics searches at the Large Hadron Collider require efficient simulations of particle propagation and interactions within the detectors. The most computationally expensive simulations involve calorimeter showers.
Advances in deep generative modelling -- particularly in the realm of high-dimensional data -- have opened the possibility of generating realistic calorimeter showers orders of magnitude more quickly than physics-based simulation.
However, the high-dimensional representation of showers belies the relative simplicity and structure of the underlying physical laws.
This phenomenon is yet another example of the \emph{manifold hypothesis} from machine learning, which states that high-dimensional data is supported on low-dimensional manifolds.
We thus propose modelling calorimeter showers first by learning their manifold structure, and then estimating the density of data across this manifold.
Learning manifold structure reduces the dimensionality of the data, which enables fast training and generation when compared with competing methods.\label{sec:abs}
\end{abstract}

\section{Introduction}
\label{sec:intro}

The accurate simulation of particle detectors is a primary component of the high energy physics program. It allows us to map the data collected by an experiment to a theoretical description of the fundamental interactions of nature.
Experimental collaborations at the Large Hadron Collider (LHC) rely on Monte Carlo simulations of physics from first-principles to emulate the passage of billions of particles through the several stages of the LHC detectors. A particularly involved step of the simulation pipeline is the generation of calorimeter showers. An incident particle interacts with an active material in the calorimeter, producing a shower of secondary particles. In turn, the shower deposits its energy into an array of scintillators which measure the energy distribution.  

Currently, LHC calorimeter shower generation is done with Geant4 \cite{geant4, geant4-add1, geant4-add2}, the state-of-the-art simulator of the passage of particles through matter. However, these simulations are very time consuming, taking up to tens of minutes per event. As the LHC enters its High Luminosity era, Monte Carlo generation will only become more complex and computationally expensive, emphasizing the need for alternative approaches enabling fast simulation.

Recent advances in the field of deep generative modelling present a solution: train a surrogate model designed to emulate calorimeter showers, using data on incident particles and their resulting showers collected from experiment or expensive simulation. Deep generative models (DGMs) aim to learn the distribution of their training data, and can generate new data by sampling from the learned distribution.
In the context of calorimeter showers, various types of generative models have been used including generative adversarial networks (GANs) \cite{goodfellow2014generative} by CaloGan \cite{paganini2018calogan}, normalizing flows (NFs) \cite{rezende2015variational, dinh2016density} by CaloFlow \cite{krause2021caloflow, krause2021caloflow2}, and score-based generative models (SGMs) \cite{song2019} by CaloScore \cite{mikuni2022score}. Each approach has aimed to improve on various aspects of the problem. GANs allow fast simulation, but not inference (the ability to evaluate the likelihood of a datapoint according to the learned distribution). SGMs provide excellent quality generated samples, but are slow to train and sample from. NFs allow inference and are trained by the seemingly principled approach of maximum-likelihood estimation. 

The main drawback of NFs is that they model a density that has the same dimensionality as the input data. For calorimeter showers, the data is presented as the energy deposited per voxel of the calorimeter which can be thought of as a vector in $\mathbb{R}^n$, where $n$ is the number of voxels. Typically, $n$ is in the hundreds or thousands, so the representation of the data is high-dimensional. This is in contrast to the relatively simple and highly structured underlying physical processes. For example, in an electromagnetic calorimeter with an incident photon, most of the interactions can be described by quantum electrodynamics \cite{fabjan2003}. 
The true distribution of electromagnetic showers is unknown, but its dimensionality is likely to be much smaller than $n$. In the machine learning context, this is an example of the \emph{manifold hypothesis}, which states that high-dimensional data clusters around a low-dimensional embedded submanifold in the ambient space \cite{bengio2013representation,pope2021}.

\begin{wrapfigure}[15]{r}{0.45\textwidth}
\vspace{-12pt}
\begin{center}
\centerline{
\includegraphics[width=0.26\textwidth,trim=20 125 80 140, clip]{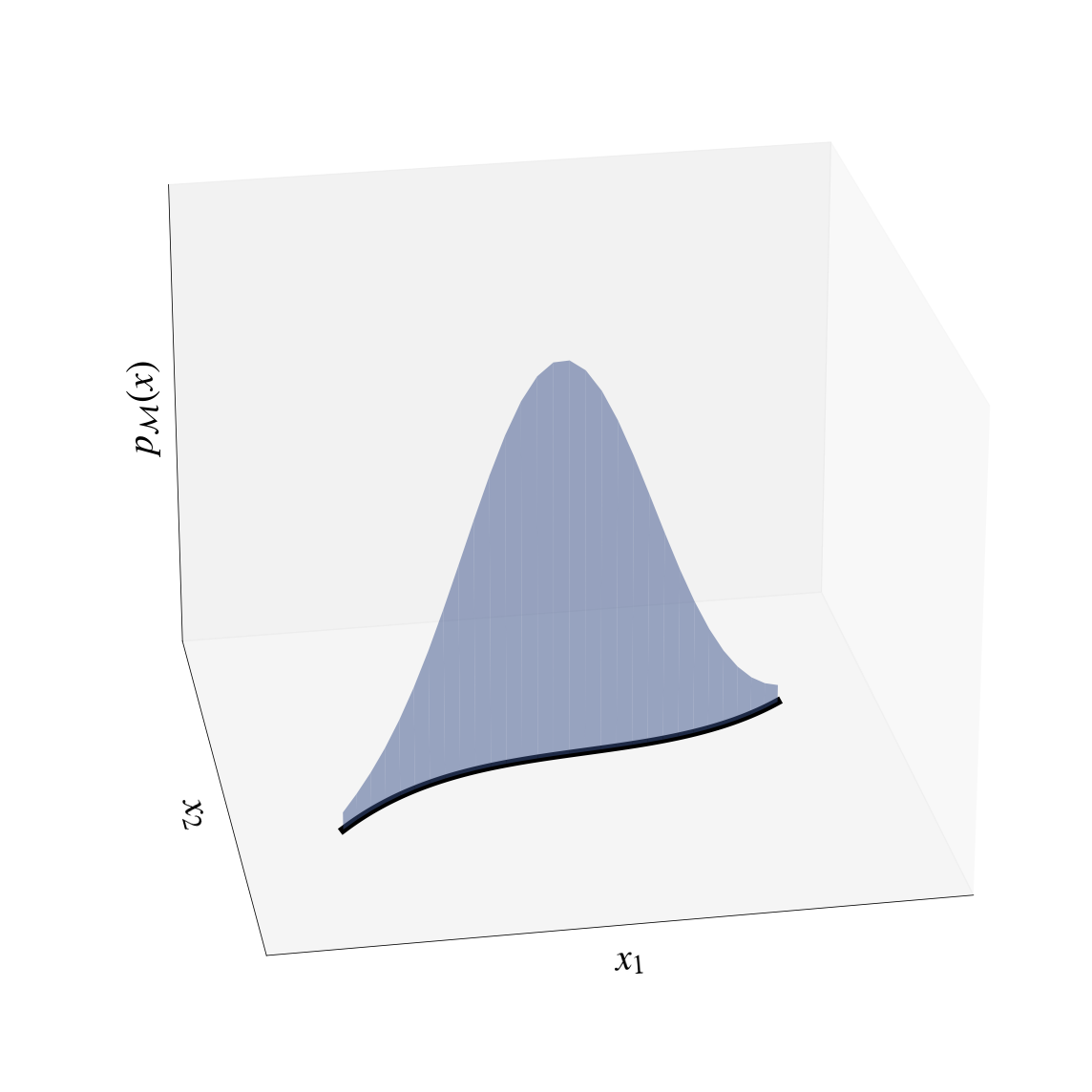}
\includegraphics[width=0.26\textwidth,trim=80 125 20 140, clip]{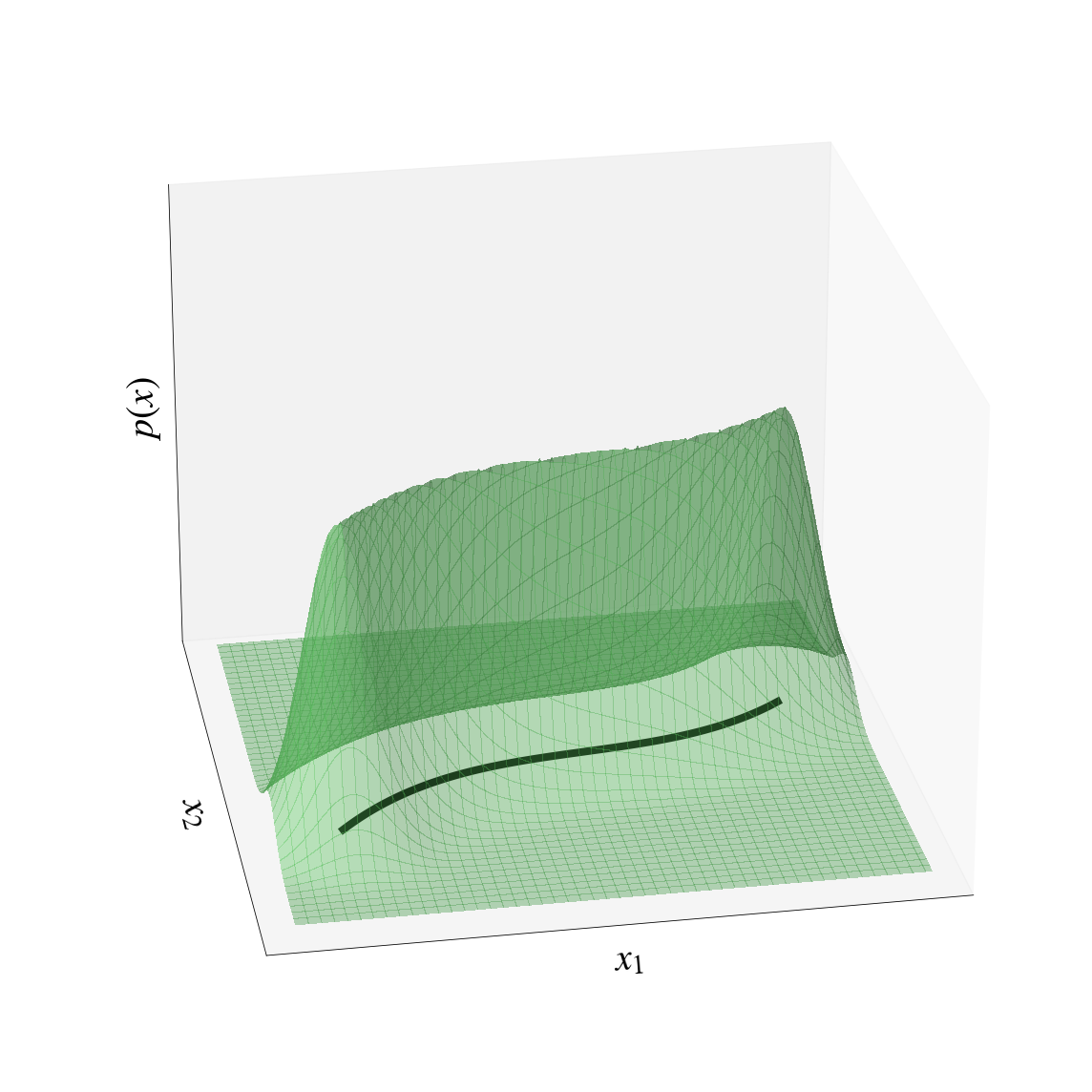}
}
\caption{A low-dimensional density on a manifold (left), and a full-dimensional density model undergoing manifold overfitting (right). Although the full-dimensional model concentrates around the manifold, it distributes the density incorrectly along the manifold.}
\label{fig:manifold_overfitting}

\end{center}
\end{wrapfigure}

Maximum-likelihood models like NFs are built on the assumption that the underlying distribution possesses a full-dimensional probability density $p(x)$ in the data space. When the data is confined to a low-dimensional manifold, this assumption is false: the data manifold is a subset of measure zero, over which no continuous density can be integrated to obtain non-zero probabilities. This mismatch between the model and the data leads to a phenomenon known as \textit{manifold overfitting} \citep{loaiza2022diagnosing}, in which the model can maximize likelihood by concentrating probability density around the manifold while incorrectly distributing the density along the manifold.
The resulting phenomenon is illustrated in Fig. \ref{fig:manifold_overfitting}. As a result, maximum-likelihood is an ill-posed objective for manifold-supported data.

A common method when training NFs and other DGMs is to add full-dimensional Gaussian noise to the training data \cite{vincent2008extracting, vincent2011connection, krause2021caloflow}. 
Adding noise can avoid the problem of manifold overfitting, but at the cost of no longer modelling the true distribution of the data, and not recovering the manifold structure \citep{horvat2021density} which could contain interesting physics.

In this work we propose to better model calorimeter showers by first learning their manifold structure and then estimating the density within \cite{loaiza2022diagnosing}. This technique allows us to avoid manifold overfitting, while the dimensionality reduction step increases the efficiency of training and shower generation. 

\label{sec:method}
\section{Method}

\subsection{Learning the Manifold of Calorimeter Showers}

While it would be interesting to understand the exact distribution of showers from first-principles, for practical applications it suffices to learn the manifold $\M$ and probability density $p_\M(x)$ from data. To do so, we follow the general two-step procedure outlined by \citet{loaiza2022diagnosing}. The first step is to learn the manifold using a \textit{generalized autoencoder}: any model that can construct low-dimensional latent encodings $z = \phi(x)$ of high-dimensional data $x$, and reconstruct the original data via an inverse transformation $x = \psi(z)$. The learned low-dimensional encodings $z$ act as coordinates on the data manifold. The class of generalized autoencoders includes the autoencoder \citep{rumelhart1985learning}, variational autoencoder \citep{kingma2013auto}, Wasserstein autoencoder \citep{tolstikhin2017wasserstein}, bidirectional GAN \citep{ donahue2016adversarial, dumoulin2016adversarially}, and adversarial variational Bayes \citep{mescheder2017adversarial}, among others. 

The second step is to perform \textit{density estimation} in the $z$ coordinates, obtaining low-dimensional densities $p(z)$. Any DGM that explicitly constructs $p(z)$ can be used, including NFs, energy based models \citep{du2019implicit}, auto-regressive models \cite{uria2013rnade}, score-based models \cite{song2019generative}, and diffusion models \cite{ho2020denoising}. Variational autoencoders and adversarial variational Bayes are also suitable as density estimators.

By capturing a probability density \textit{within} the manifold, the two-step procedure evades the dimensionality mismatch in maximum-likelihood estimation. If required, probability densities in the data space can be computed using a change of metric from the low-dimensional coordinates into the high-dimensional space \cite{gemici2016normalizing, caterini2021rectangular, ross2021tractable}:
\begin{equation}
    p_\M(x) = p(z)\det \left(J_\psi(z)^T J_\psi(z)\right)^{-\tfrac{1}{2}},
\end{equation}
where $J_\psi(z)$ is the Jacobian of $\psi$ evaluated at $z = \phi(x)$. Since the computation of Jacobians can be expensive, we emphasize that maximizing $p(z)$ directly is sufficient for training. In addition to being more mathematically principled, two-step models are highly performant: stable diffusion \cite{rombach2021highresolution} is one example in which a diffusion model is trained on the data manifold learned by an autoencoder, and is capable of surprisingly photorealistic image generation.
Calorimeter shower simulation requires a model that can simulate showers conditional on incident energies. We assume the manifold contains showers $x$ corresponding to all possible incident energies $E_{\text{inc}}$ and aim to model the conditional density $p_\M(x\mid E_{\text{inc}})$. The generalized autoencoder only needs to map showers between the data space and latent space, which does not require knowledge of $E_{\text{inc}}$, hence we only add conditioning to the density estimator. When $E_{\text{inc}}$ information is provided to the generalized autoencoder it can more easily cluster the shower data, leading to a segmented latent representation that may not generalize to unseen $E_{\text{inc}}$.

\subsection{Intrinsic Dimension Estimation} 
In most examples of generalized autoencoders the dimensionality $d$ of the latent representation is not learned, but is specified as a hyperparameter of the method. If $d$ is not known \emph{a priori}, it could be set empirically by trying many values and selecting the one with optimal validation performance, but training many models is computationally expensive. 
Instead, we employ a statistical estimator of the intrinsic dimension, several of which were reviewed in \cite{pope2021} and used to provide evidence for the manifold hypothesis. Here we select the Levina-Bickel estimator \cite{levina2004} with the MacKay-Ghahramani correction \cite{mackay2005} because it is efficient to compute, and correlates well with synthetic datasets of known dimensionality \citep{pope2021}. The estimator is
\begin{equation}\label{eq:estimator}
    \hat{d_k} = \left( \frac{1}{n(k-1)} \sum_{i=1}^n \sum_{j=1}^{k-1} \log \frac{T_k(x_i)}{T_j(x_i)}\right)^{-1},
\end{equation}
where $T_k(x_i)$ is the Euclidean distance between datapoint $x_i$ and its $k$th nearest neighbour in the dataset $\{x_i\}_{i=1}^n$. The hyperparameter $k$ sets the scale at which the manifold is probed, with smaller values giving a more close up view. In the following we estimate $\hat{d_k}$ on the training data, then use it to specify the generalized autoencoder architecture.

\label{sec:experiments}
\section{Experiments}
\subsection{Dataset}\label{sec:dataset}

As a proof of concept, we model the photon dataset provided by the Fast Calorimeter Simulation Challenge 2022 \cite{calochallenge}. The training and test datasets, each containing 121,000 electromagnetic calorimeter showers simulated by Geant4, display a cylindrical geometry divided into five layers of bins in polar coordinates, with 368 voxels of deposited energies. The incident photons are parallel to the $z$-axis of the calorimeter, and have energies ranging from 256 MeV to over 4 TeV, increasing in powers of two, with each level represented by 10,000 samples, except at higher energies where there are fewer. We perform additional preprocessing to the dataset to facilitate training, similar to \cite{krause2021caloflow} and \cite{mikuni2022score}. Due to noise, the total deposited energy $E_{\text{dep}}$ can be greater than the incident energy $E_{\text{inc}}$, with the largest ratio $E_{\text{dep}}/E_{\text{inc}}$ in the training set as $r_{\text{max}}=3.1$. Each voxel of each shower is rescaled by $r_{\text{max}}E_{\text{inc}}$ to ensure voxels are in $[0,1]$. We append $E_{\text{dep}}$ as an additional feature, and linearly scale incident energies into the range $[0,1]$. The training dataset is split 80/20 for validation.

%%%% Fix awkward page break
\newpage

\subsection{Intrinsic Dimension Estimates}\label{sec:dim}

\begin{wrapfigure}[15]{r}{0.4\textwidth}
\vspace{-14pt}
\centering
    \includegraphics[width=0.4\textwidth, trim={5 8 8 5}, clip]{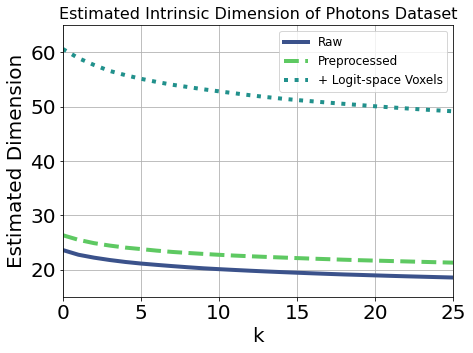}
    \caption{The estimated intrinsic dimension of the photon training dataset as a function of the hyperparameter $k$. }
    \label{fig:dim_estimation}
\end{wrapfigure} 
We apply the dimension estimator Eq. \eqref{eq:estimator} to the photon training dataset before and after preprocessing, with results shown in Fig. \ref{fig:dim_estimation}. Other works \cite{krause2021caloflow, mikuni2022score} transform voxels to logit space which has a large effect on the estimated dimension since distances between datapoints are stretched. The logit transformation may have had empirical benefits in these prior works in part because it increased the ``effective'' intrinsic dimension and thereby reduced manifold overfitting in models that had mismatched dimensionality. Our model does not have this issue so we do not use the logit transformation.
At very small $k$ the estimator is biased to overestimate \cite{levina2004, mackay2005}. Larger $k$ means more distant neighbours are considered for each point, effectively looking at longer length scales which can smooth out noise but also miss small (compact) dimensions. This intuition explains the decreasing nature of $\hat{d_k}$ with $k$. Based on the value at $k=10$ which has provided an accurate estimate in prior work \citep{pope2021, brown2022union}, we use the dimension $\hat{d}_{10}=20$ for our experiments. 

\subsection{Manifold Learning and Density Estimation}

\begin{figure}[t]
\centering
  \includegraphics[width=0.48\textwidth, trim={0 0 0 0}, clip]{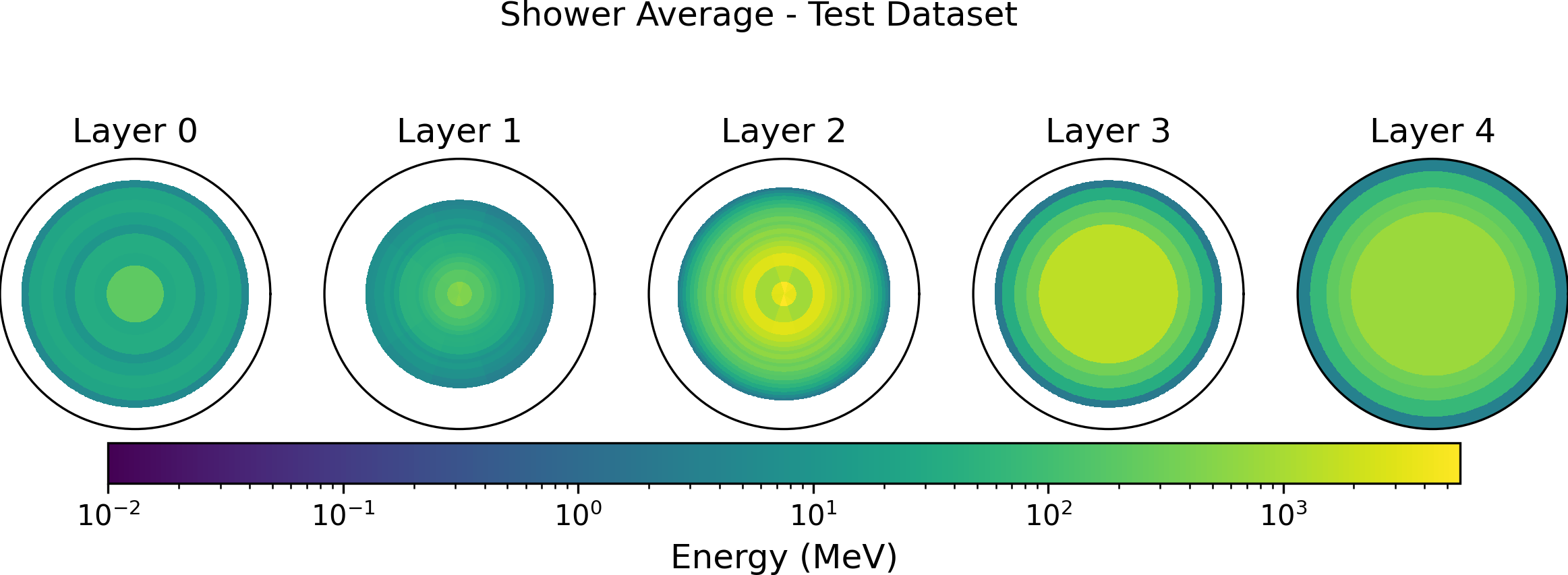} \quad 
  \includegraphics[width=0.48\textwidth, trim={0 0 0 0}, clip]{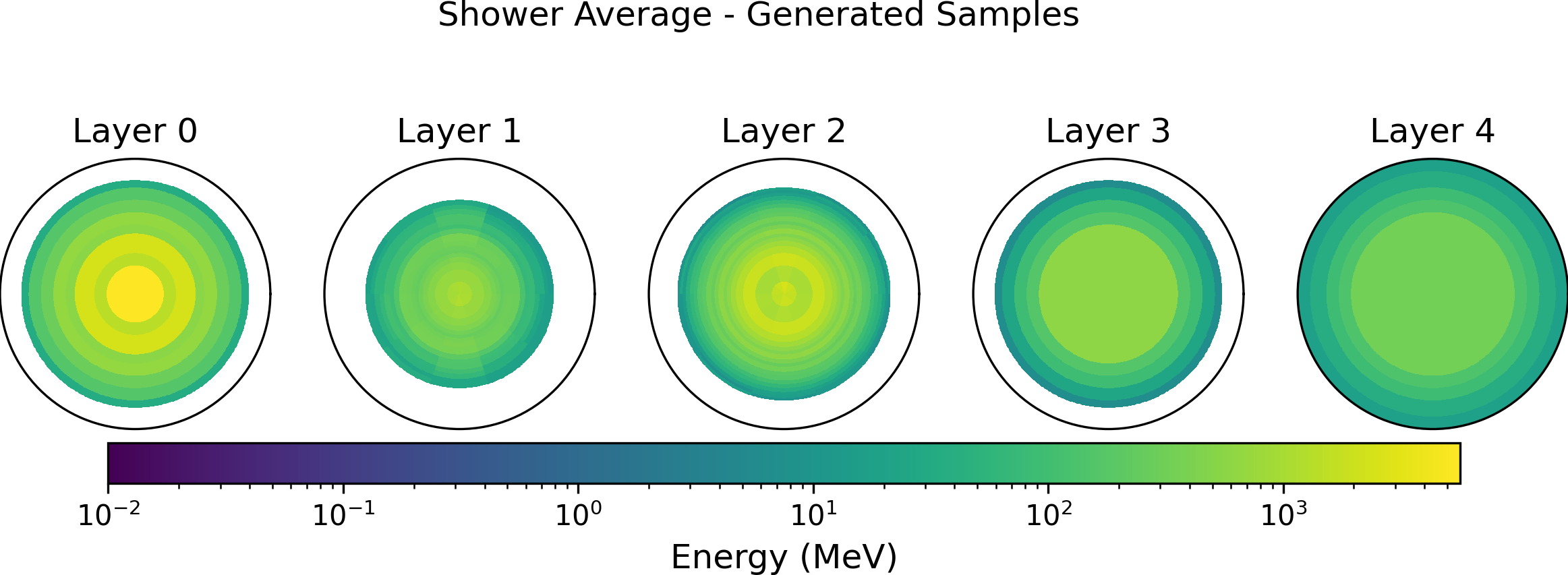}
    \caption{Average deposited energy per voxel for the test dataset (left), and generated samples (right). The samples were conditionally generated using the distribution of energies from the test set.}
    \label{fig:average_showers}
    \vspace{-10pt}
\end{figure}

\begin{wrapfigure}[18]{r}{0.35\textwidth}
\vspace{-16pt}
\centering

\small
  \makeatletter\def\@captype{table}\makeatother% "Change float to table"
\caption{Histogram $\chi^2$ separation powers in high-level features. Lower is better. L denotes layer. CE is the center of energy.}\label{tbl:hist}
\centering
    \scalebox{0.9}{
    \begin{tabular}{rr l}
        \toprule
        \textsc{Feature} & $\chi^2$ \textsc{Power} \\
        \midrule
        $E_{\text{dep}}/ E_{\text{inc}}$ & 0.0535  \\ 
        $E_{\text{dep}}$, \textsc{L}$0$ & 0.0540  \\ 
        $E_{\text{dep}}$, \textsc{L}$1$ & 0.0304  \\ 
        $E_{\text{dep}}$, \textsc{L}$2$ & 0.0243  \\ 
        $E_{\text{dep}}$, \textsc{L}$3$ & 0.0045  \\ 
        $E_{\text{dep}}$, \textsc{L}$4$ & 0.0009  \\ 
        CE in $\eta$, L1 & 0.0376  \\ 
        CE in $\eta$, L2 & 0.0512  \\ 
        CE in $\phi$, L1 & 0.0145  \\ 
        CE in $\phi$, L2 & 0.0391  \\ 
        Width in $\eta$, L1 & 0.1548  \\ 
        Width in $\eta$, L2 & 0.0538  \\ 
        Width in $\phi$, L1 & 0.1080  \\ 
        Width in $\phi$, L2 & 0.0489  \\ 
        \bottomrule
    \end{tabular}
    }
\end{wrapfigure}

We modelled showers with a VAE as our generalized autoencoder and a conditional NF for density estimation on the learned manifold, using the code of \citet{loaiza2022diagnosing}. The VAE's encoder and decoder architectures were multi-layer perceptrons with 3 hidden layers of 512 units.\footnote{Full experimental details are provided in App. \ref{app:A}.} The NF, implemented with the code of \citet{nflows}, consisted of a 4-layer rational-quadratic neural spline flow \cite{durkan2019neural} and a 3-block residual network \cite{he2016deep} at each layer. The output of each residual block was combined with the conditioning input using a gated linear unit \cite{dauphin2017language}. The VAE and NF were each trained for 200 epochs for a total runtime of 110 minutes on a Titan V GPU, requiring only 1 GB of memory. In comparison, CaloScore \cite{mikuni2022score} used 16 A100 GPUs and trained for 5 times as many epochs.

In Fig. \ref{fig:average_showers} we show the average deposited energy per voxel for the test dataset, and for conditional samples from our model using the same incident energy distribution as the test set. The rotational symmetry of layers 1 and 2 was learned well, but our model allocated too much energy to layer 0. Conditioning the generalized autoencoder on incident energies could improve this.

\begin{figure}[t]
    \centering
        \includegraphics[width=0.23\textwidth]{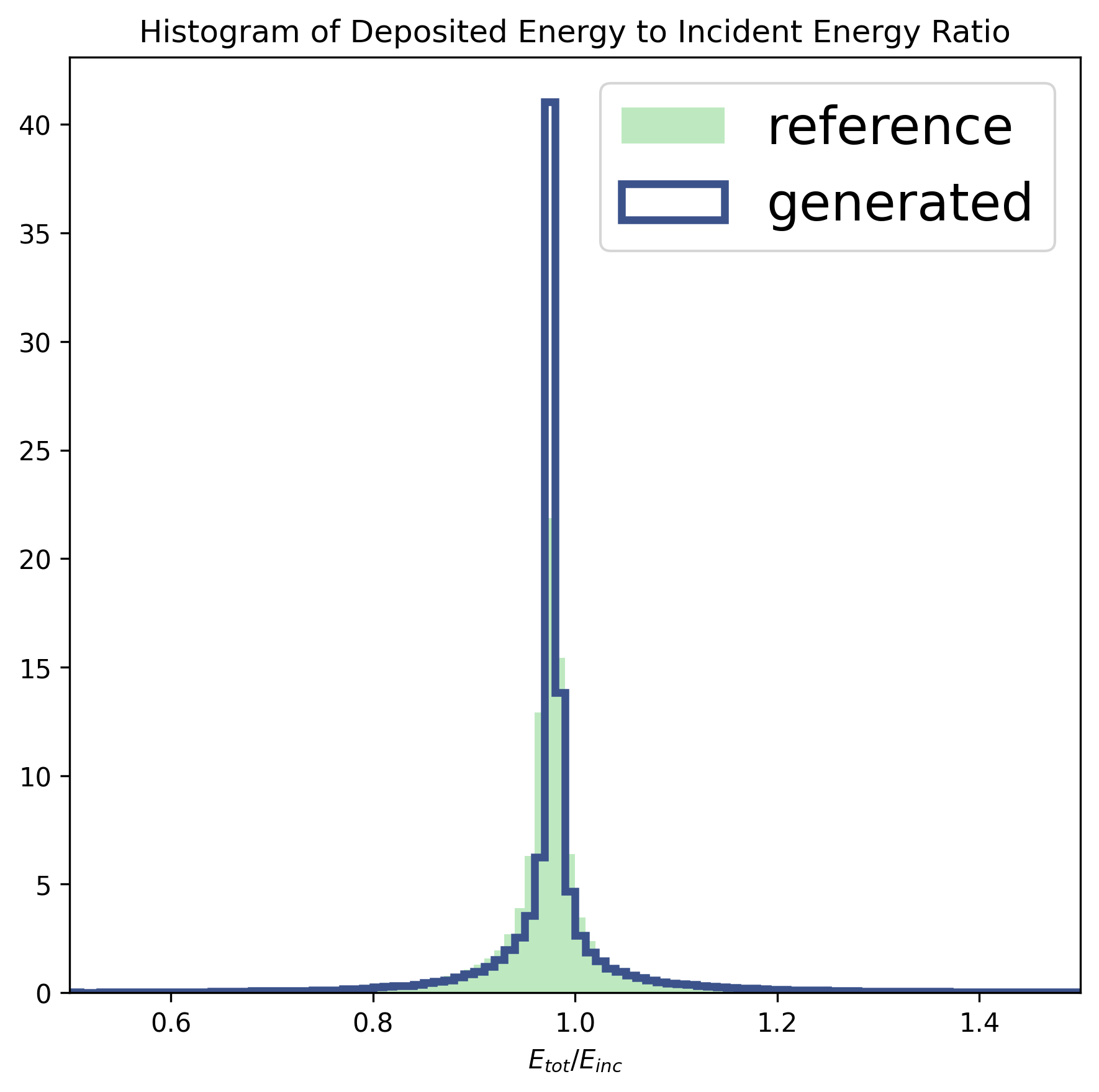}
        \includegraphics[width=0.23\textwidth]{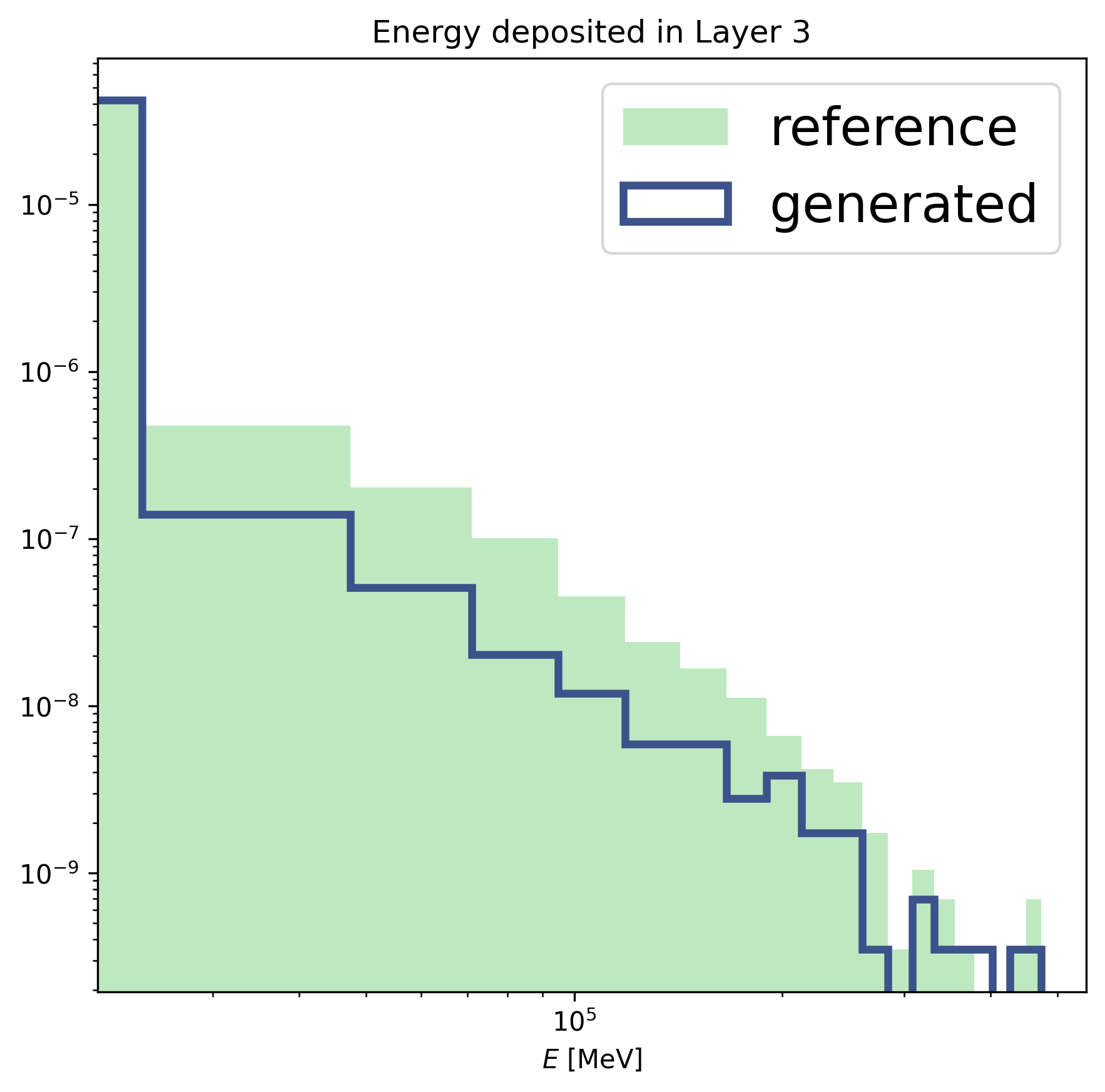}
        \includegraphics[width=0.23\textwidth]{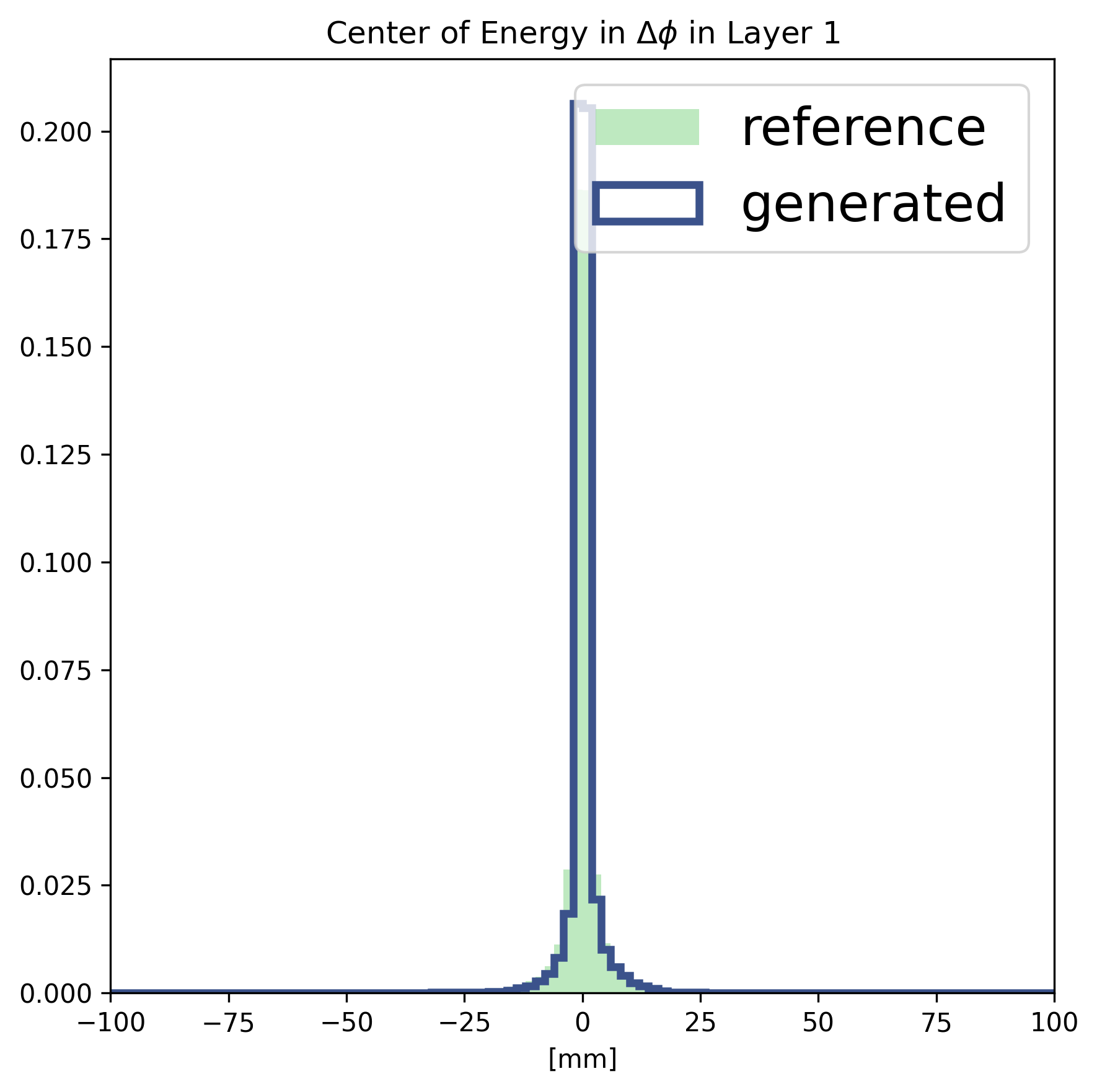}
        \includegraphics[width=0.23\textwidth]{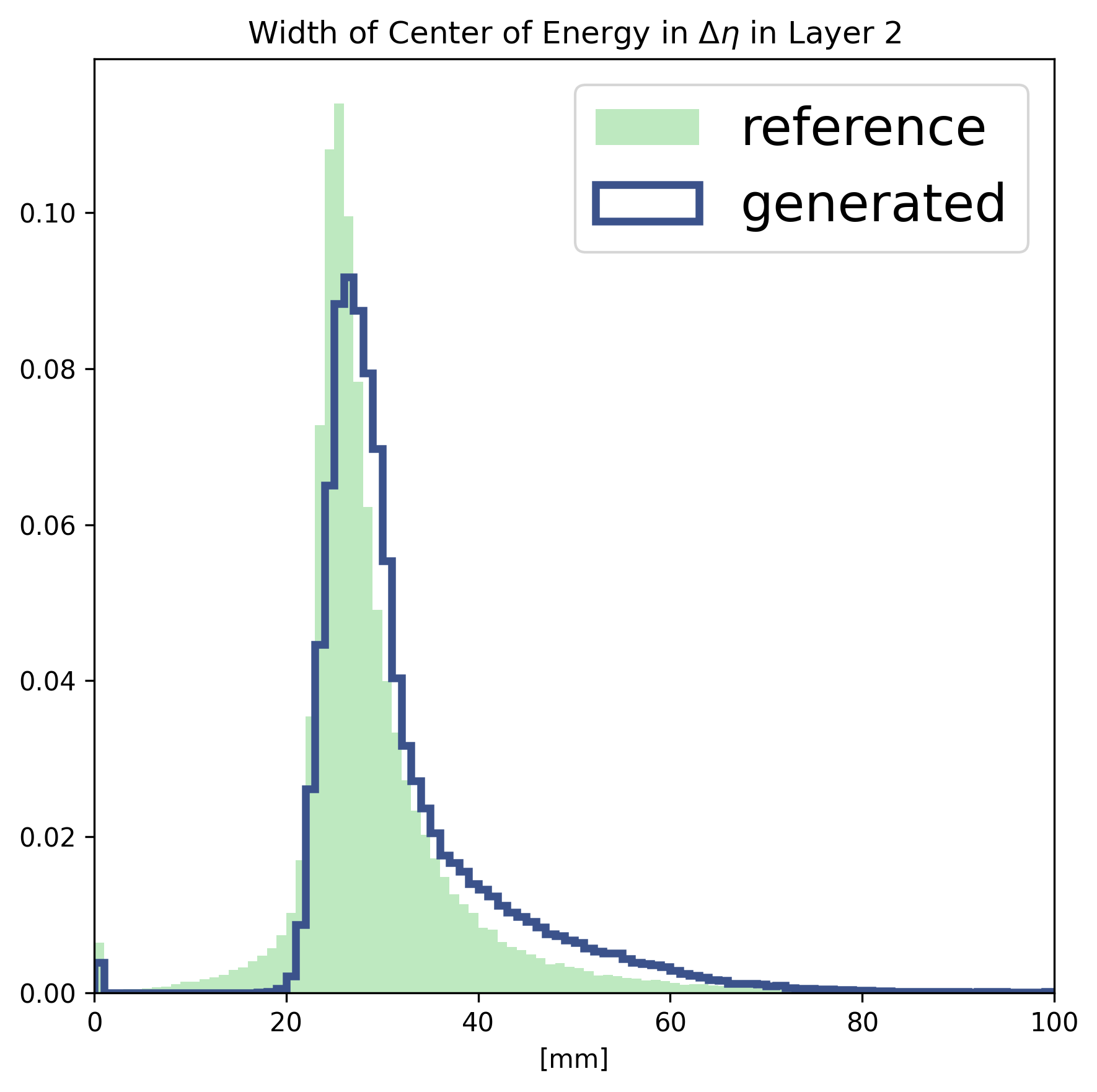}
    \caption{Histograms of high-level features comparing generated samples to the test set.}
    \label{fig:histograms}
    \vspace{-15pt}
\end{figure}

In Table \ref{tbl:hist} we compare our model's samples to the test set using histograms of high-level features as described in the Challenge \cite{calochallenge}. For each feature the $\chi^2$ separation power between histograms is computed, with lower numbers indicating the model matching the true distribution. Deposited energies and the centers of energy are well-learned, with some improvement required for widths. Four corresponding histograms are shown in 
Fig. \ref{fig:histograms}. Using the architecture and training details provided in \cite{krause2021caloflow}, we trained a binary classifier to distinguish real and generated showers which achieved only 0.78 AUC. An AUC lower than 1.0 shows that many generated showers are indistinguishable from real ones. By comparison, CaloScore \cite{mikuni2022score} reported 0.98 AUC.

\begin{wrapfigure}[9]{r}{0.35\textwidth}
\vspace{-20pt}
\centering
\small
    \setlength{\tabcolsep}{2.6pt}

      \makeatletter\def\@captype{table}\makeatother% "Change float to table"
\caption{Sample generation times}\label{tbl:times}    
\scalebox{0.9}{
    \begin{tabular}{rrr}
        \toprule
        \shortstack{\textsc{Batch}\\ \textsc{Size}} & \shortstack{\textsc{Number of}\\\textsc{showers}} & \shortstack{\textsc{Time per}\\\textsc{shower} (ms)} \\
        \midrule
        1,000 & 1,000 & 0.0598  \\
        1,000 & 100,000 & 0.0844  \\
        10,000 & 10,000 & 0.0265  \\
        10,000 & 100,000 & 0.0246  \\
        50,000 & 50,000 & 0.0216  \\
        50,000 & 100,000 & 0.0201  \\
        \bottomrule
    \end{tabular}
    }
\end{wrapfigure}
Our method also demonstrates a significant speedup in generation time compared to others. In Table \ref{tbl:times} we show the amortized time to generate each sample for various batch sizes and numbers of samples generated. We include both the runtime of the model, and overhead to undo preprocessing on the samples which is required to produce showers in their original format. For large batch sizes, which still comfortably fit on a single Titan V GPU, 
our generation time is as low as 0.02 ms per shower. In comparison, CaloScore \cite{mikuni2022score} reported a sampling time of 40 ms per shower on the same dataset, while CaloFlow II's best settings (on a similar dataset) required 0.08 ms per sample \cite{krause2021caloflow2}. The Geant4 simulation times increase with incident energy, and range from 100 ms to 3 s \cite{aad2021atlfast3}.

\label{sec:discussion}
\section{Discussion}
In this work we investigated deep generative models for generating calorimeter showers that learn a low-dimensional manifold structure and estimate densities on it. While the calorimeter data was represented in 368-dimensional space, we found the intrinsic dimension of the shower manifold to be approximately 20, and showed that it is possible to accurately learn the data's density in so few dimensions. Compared to previous approaches, our use of dimension reduction led to lightweight models, significant speedups in training and sample generation, and good performance due to the avoidance of \emph{manifold overfitting} \cite{loaiza2022diagnosing}. While this work is preliminary, we expect that further improvements are possible as the methods are scaled up.

For calorimeter shower simulation, past work \cite{paganini2018calogan, krause2021caloflow, aad2021atlfast3, mikuni2022score} has focused solely on creating fast and accurate density models, and has achieved these goals to varying extents. However, past models do not inform us about the nature of the manifold supporting the data distribution, which could reveal interesting physics. Furthermore, modelling mismatches could impose fundamental barriers to accurate simulation. Consider the topology of the learned distribution - normalizing flows learn a diffeomorphism that acts on a simple prior distribution and hence will be unable to learn topologically non-trivial data manifolds \cite{cornish2020}. In contrast, learning the manifold implicitly \cite{ross2022neural} can enrich the class of topologies that can be modelled without prior knowledge or additional assumptions. We expect that models which can better incorporate aspects like dimensionality, topology, geometry, compactness, and connectedness will not only perform better, but will provide more insight into the physical processes at play, especially when trained on experimental rather than simulated data.

% \newpage
\paragraph{Broader Impact}
While generative models certainly have far-reaching societal implications, for instance when used maliciously to generate images of humans  \cite{brundage2021malicious}, or new chemical structures which could be weaponized \cite{urbina2022dual}, we do not foresee negative societal outcomes from applying them to calorimeter showers. It should be acknowledged that the utility of surrogate models is limited by the quality of data they are trained on. In this setting, our method relies on accurate simulations of calorimeter showers from Geant4. Our method is not meant to replace physics-based simulation, but complement it by addressing the growing concern of computational cost.

\begin{ack}
M.L. acknowledges the financial support of the European Research Council (grant SLING 819789).  The research of H.R.G. is funded by the Italian PRIN grant 20172LNEEZ.
\end{ack}

{
\small
\bibliography{bib}
\bibliographystyle{abbrvnat}
}

% \newpage
\appendix
\section{Appendix}\label{app:A}

In this appendix we provide complete details on our model architectures, experimental setup, and evaluation approach.

\subsection{Model Architectures}

For the experiments on the photon dataset we used a VAE as the generalized autoencoder to learn a representation of the manifold, and then trained a NF on the latent representation to estimate the density. As described in Sec.  \ref{sec:dim}, the latent space dimension was fixed at 20 based on our intrinsic dimension estimates of the data manifold.

The VAE's encoder and decoder were both multi-layer perceptron neural networks with three hidden layers of 512 units each, and ReLU activations. The encoder output the means and variances for a diagonal Gaussian over the latent dimensions. The decoder output was also treated as a diagonal Gaussian with means for each data dimension but only a single variance shared across all dimensions. The prior distribution was a unit variance diagonal Gaussian over the latent space.

The NF was a 4-layer rational-quadratic neural spline flow with a 3-block residual network in each layer. The output of each residual block was combined with the conditioning input, namely the incident energy, using a gated linear unit. The NF's prior distribution was a unit variance diagonal Gaussian.

\subsection{Experiment details}

All models were trained with batch sizes of 512 showers, the Adam optimizer \cite{kingma2014adam} with a learning rate of 0.001, and gradient clipping with a max gradient norm of 10.

The VAE and NF were trained sequentially. In the first step, the VAE was trained for 200 epochs with cosine annealing from the initial learning rate. Once the VAE was trained, its parameters were frozen and the training dataset was encoded deterministically with the VAE encoder means. Then the data was rescaled by the absolute maximum element value of any datapoint. Finally, the NF was trained for 200 epochs with early stopping after 100 epochs of no validation improvement. The validation metric was the average $\chi^2$ separation power over all high level features, see Table \ref{tbl:hist}, using 50,000 generated samples for the comparison. The epoch with the best validation metric was used for evaluation.

\subsection{Evaluation}

We evaluated our two-step model by visually comparing average showers, computing histogram separation powers for high-level features, and by training a binary classifier to distinguish real and generated showers. Each method requires samples from the model which were generated by conditionally sampling from the NF prior, passing through the NF, decoding with the VAE decoder, and finally undoing the preprocessing steps described in Sec. \ref{sec:dataset}. The decoding was deterministic, only using the mean of the VAE output distribution. Showers were sampled conditionally using the same number of showers and distribution of incident energies as the test set.

The binary classifier was trained to distinguish test dataset points from a fixed set of generated samples as described above. The test dataset and generated samples were separately split in a 60/20/20 ratio and respectively joined to form classifier training, validation, and test sets. The classifier itself was a multi-layer perceptron with three hidden layers of 512 units each, and ReLU activations. It was trained with the Adam optimizer and a learning rate of 0.0005 for a maximum of 60 epochs with early stopping after 10 rounds of no improvement in accuracy on the validation set. The epoch with best validation accuracy was used to finally evaluate the classifier's AUC on its test set.

\end{document}